\documentclass[11pt]{article}
\usepackage{hyperref}
\usepackage{amsmath}
\usepackage{amssymb}
\usepackage{bm}
\usepackage{latexsym}
\usepackage{graphicx}
\usepackage{euscript} 
\usepackage{mathrsfs}
\begin{document}
\title{Imprints of intrinsic and exterior curvatures in cosmology}
\author{Ali A. Asgari \\Amir H. Abbassi*\\
\\\vspace{6pt} Department of Physics, School of Sciences,\\ Tarbiat Modares University, P.O.Box 14155-4838, Tehran, Iran\\}
\maketitle
\begin{abstract}
Nowadays, according to the observational evidences the curvature parameter of the universe is neglected and spatially flat FLRW model is on the top of interest for cosmologists. However, due to some discrepancies between  $ \Lambda $-CDM model anticipations and observations, one may think out the curvature parameter as the solution even though it may be very small. So, in this article we investigate the eras in which the curvature influence was or is more significant. In addition, a geometrical interpretation of the Hubble parameter is dedicated. We find that the Hubble parameter is more appropriate  to be defined as the mean curvature of the spatial section of the universe because the concept of the scale factor in a nonhomogeneous universe is not precisely clear. Moreover, we realize that by employing the linear perturbations, two principle curvatures of the absolute space change into what depends on the local Hubble parameter. 
\end{abstract}
\section{Introduction}
\label{intro}
Among the parameters describing $ \Lambda $-CDM model, Hubble constant $ H_0 $ has an exclusive situation as the sign of expanding universe. Suppose $ a $  is the scale factor of the universe, then the Hubble parameter is defined as $ H=\frac{\dot{a}}{a} $ which dot stands for the derivative respect to the cosmic time. The value of $ H $ at present time is called the Hubble constant which is denoted by $ H_0 $ . According to the last reported result by Planck collaborations $ H_0=67.4\pm0.5\,\rm{KmMPc}^{-1}s^{-1} $ \cite{ref1} although new local measurements indicate different value $ H_0=73.3^{+1.7}_{-1.8}\,\rm{KmMPc}^{-1}s^{-1} $ \cite{ref2}. The value of $ H_0 $ reflects the rate of universe expansion. Moreover, the age of the universe is proportional to the Hubble time$ \frac{1}{H_0} $\cite{ref3} ($ \frac{1}{H_0} $ sometimes is called the Hubble age of the universe). Geometrically, the Hubble parameter implies expansion rate of the spatial section of the FLRW universe. However, the FLRW model is just an approximation which ignores many things in the universe such as galaxy clusters, galaxies, stars etc. More realistically, we may invoke linear perturbation theory to describe departures from homogeneity and isotropy. So, what is the meaning of the Hubble constant now? In this paper, we investigate geometric meaning of the Hubble parameter in the perturbed FLRW universe. We survey the intrinsic and exterior properties of the spatial section of the perturbed FLRW universe too. It shall be cleared that the exterior curvatures of the spatial slice against the intrinsic curvature don't depend on the curvature index $ K $. Furthermore, we discuss about the eras in when the curvature has more remarkable imprint.   
\section{Effect of the curvature parameter in various cosmological eras: a simple analysis}
\label{sec:1}
Now we are going to recognize eras in which the curvature is more influential. First notice that the \textit{sectional} (\textit{Riemannian}) \textit{curvature} of the spatial section of the universe at $ t $ is
\begin{equation}
\mathcal{K}_s\left(t \right)=\frac{K}{a^2\left( t\right)}=-H^2\Omega_K\left( t\right) . 
\end{equation}
Now let's define the \textit{curvature density} as
\begin{equation}
\rho _K\left( t\right):=-\frac{3}{8\pi G}\mathcal{K}_s\left(t \right).
\end{equation}
Consequently
\begin{displaymath}
\Omega_K\left( t\right)=\frac{\rho_K\left( t\right)}{\rho_{crit}\left( t\right)}=-\frac{K}{H^2a^2}.
\end{displaymath}
Besides, for a barotropic cosmic fluid described with equation of state $  p=\omega\rho $ we have \cite{ref3}
\begin{equation}
\rho\left( t\right)\propto a^{-3\omega-3}.
\end{equation}
It means that effect of curvature in the universe is equivalent to a barotropic cosmic fluid with $ \omega=-\frac{1}{3} $, so every when in the early stage for the cosmic fluid $ \omega <-\frac{1}{3} $ i.e. the strong energy condition is contradicted, the universe would be \textit{curvature-dominated}. Certainly, in these eras the universe had been accelerated. So, the inflationary era for which $ \omega\simeq -1 $ is curvature-dominated \cite{ref4}. It means the curvature -if it exists- has played a crucial role in the dynamics of the very early universe. In other words, the nature of the spatially curved inflationary universe would be basically different from spatially flat case. For example Ellis and his collaborators \cite{ref5,ref6} analyzed the spatially closed inflationary model and revealed the causal and dynamical discrepancies between spatially flat and closed models. Furthermore, the spectra coming out from spatially closed inflationary model is different from ordinary inflationary scenario \cite{ref7,ref8}. Similar treatment is observed in spatially open model \cite{ref9,ref10}. \\
It is possible to analyze importance of curvature relative to the perturbations scales. It is clear that when perturbations leave the horizon, the curvature role is negligible. For $ K=+1 $ model, suppose that number of e-foldings from $ t=0 $ to the time for which the perturbation with scale $ \frac{1}{H_0 a_0} $ leaves the horizon is \cite{ref4}
\begin{equation}
\Delta\mathcal{N}\left(0\rightarrow t_\ast \right)=\ln\cosh\frac{t_\ast}{\alpha}.
\end{equation}
Besides 
\begin{equation}
\Omega_K =-\frac{1}{\dot{a}^2\left(t_\ast \right) }=-\frac{1}{\sinh^2\frac{t_\ast}{\alpha}}.
\end{equation}
Consequently, the number of e-foldings till the largest observable scale in the universe at $ t_0 $ leaves the horizon is
\begin{equation}
\Delta\mathcal{N}\left(0\rightarrow t_\ast \right)=\frac{1}{2}\ln\left(1-\frac{1}{\Omega_K} \right).
\end{equation}
It should be noticed that more inflation i.e. more number of e-foldings leads to a universe which is more spatially flat\cite{ref5}. Besides appealing role of the curvature on dynamics of the inflationary epoch, its influence at post-inflationary eras is not deniable. For example, in \cite{ref11} imprints of curvature parameter on evolution of cosmological spectral indices after inflation has been considered exhaustively. 
\section{The Hubble parameter as an exterior property of the spatial slice of the universe}
Let's suppose that during the most of the time, departures from homogeneity and isotropy have been very small, so they can be treated as first order perturbations and total metric of the universe is
\begin{equation}
g_{\mu \nu }  = \bar {g}_{\mu \nu }  + h_{\mu \nu } ,\qquad h_{\mu \nu }\ll 1
\end{equation}
where $  \bar{g}_{\mu \nu } $  and $ h_{\mu \nu }  $ are the unperturbed metric and the first order perturbation respectively. Note that  $  \bar{g}_{\mu \nu } $ is the FLRW metric which in the comoving quasi-Cartesian coordinates can be written as\cite{ref3}
\begin{eqnarray*}
&&\bar{g}_{00}  =  - 1 ,\quad \bar{g}_{0i}  = \bar{g}_{i0}  = 0 ,\\
&&\bar{g}_{ij}  = a^2 \left( t \right)\tilde g_{ij}=a^2 \left( t \right)\left(\delta _{ij}  + \frac{Kx^i x^j}{1 - K{\bf{x}}^2 }\right) .
\end{eqnarray*}
Bar over any quantity denotes its unperturbed value. On the other hand, $ h_{\mu \nu } $ can be decomposed into four scalars, two divergenceless, spatial vectors and a symmetric, traceless and divergenceless spatial tensor as follows\cite{ref12,ref13,ref14,ref15}  
\begin{eqnarray}
&&h_{00}  =  - E ,\\
&&h_{i0}  = a\left( {\nabla _i F + G_i } \right) ,\\
&&h_{ij}  = a^2 \left( {A\tilde g_{ij}  + H_{ij} B + \nabla _i C_j  + \nabla _j C_i  + D_{ij} } \right)\label{SP} ,
\end{eqnarray}
where $ \nabla_{i} $ is the covariant derivative respect to the spatial unperturbed metric $\bar g_{ij}(=a^2\tilde g_{ij})$ and $ H_{ij}=\nabla _i \nabla _j $ is the \textit{covariant Hessian}. All the perturbations $A, B, E, F, C_{i}, G_{i}$ and $D_{ij}$ are functions of $t$ and $\mathbf{x} $ which satisfy 
\begin{eqnarray*}
 &&\nabla ^i C_i  = \nabla ^i G_i  = 0 , \\ 
 &&\tilde g^{ij} D_{ij}  = 0,\quad \nabla ^i D_{ij}  = 0,\quad D_{ij}  = D_{ji} . 
 \end{eqnarray*}
Notice that $ D_{ij} $ denotes the radiative part of perturbations. Now let's foliate the spacetime by considering set of slices $ t=const $ as the leaves (foliums). These leaves indicate absolute spaces which are labeled by absolute time $ t $, i.e. in every moment a new absolute space would be created. It is easy to show that the normal vector field associated with absolute spaces is 
\begin{equation}
{n^\mu } = \left(1 - \frac{E}{2} \right){\delta ^\mu }_0 - a\bar{g}^{\mu i}\left(\partial _i F + G_i\right),
\end{equation}
which corresponds to a timelike hypersurface-orthonormal congruence with
\begin{eqnarray*}
&&\Theta  = 3H - \frac{3}{2}HE + \frac{3}{2}\dot A - a \nabla ^2\sigma ,\\ 
&&\omega _{\mu \nu} \equiv 0 ,
\end{eqnarray*}
and
\begin{equation*}
\left\{
\begin{aligned}
&\sigma _{00} = \sigma _{0i} = \sigma _{i0} = 0 ,\hfill \\
&\sigma _{ij} =  - a\left( H_{ij} - \frac{1}{3}\bar{g}_{ij}\nabla ^2 \right)\sigma + \frac{a^2}{2}\left(\nabla _i \dot{C}_j +\nabla _j \dot{C}_i \right)\\ &\hspace{.9cm} -  \frac{a}{2}\left( \nabla _i G_j+ \nabla _j G_i \right)+ \frac{1}{2}a^2 \dot{D}_{ij} ,
\end{aligned}
\right.
\end{equation*}
Here $ \Theta $, $ \omega _{\mu \nu} $ and $ \sigma _{\mu \nu} $ are respectively expansion scalar, vorticity tensor and shear tensor of the congruence. Moreover $ \sigma=F-\frac{a}{2}\dot{B} $ is known as \textit{shear potential} associated with the congruence. Notice that $ \nabla ^2  = \bar{g}^{ij}H_{ij} $  is the \textit{Laplace-Beltrami operator} respect to the $\bar{g}_{ij} $. By neglecting the perturbations, one can see $ H=\frac{1}{3}\Theta $, so we may define the \textit{local Hubble parameter} as \cite{ref16}
\begin{equation}\label{H}
H\left(t,\mathbf{x}\right) := \frac{1}{3}\Theta \left( t,\mathbf{x} \right) = H - \frac{1}{2}HE + \frac{1}{2}\dot{A} - \frac{1}{3}a\nabla ^2 \sigma ,
\end{equation}
which is obviously gauge-dependent scalar. Thus the Hubble parameter is indeed an \textit{scalar random field}  over spatial slice of the universe (just like as scalars $ E $, $ A $ etc.) with mean $ H\left(t \right)  $ i.e.
\begin{equation}
\langle H\left( t, \mathbf{x}\right) \rangle  =H\left( t\right).
\end{equation}
In other words, what is measured as the Hubble constant is the mean of the stochastic field $ H\left( t, \mathbf{x}\right) $.\\
Now let's investigate the induced and intrinsic properties of the absolute spaces $ t=const $. In general, for every submanifold we may define two kinds of intrinsic and exterior curvatures \cite{ref17,ref18}. An intrinsic curvature is invariant under isometries \cite{ref18}. In other words, against extrinsic curvatures, an intrinsic curvature of a submanifold doesn't depend on choosing of immersion which embedding the submanifold to the main manifold \cite{ref18}. It can be relate exterior and intrinsic curvatures via the \textit{Gauss-Codazzi equations} \cite{ref19}. The Ricci (scalar) curvature is an intrinsic curvature for $ t=const $ ,
\begin{equation}
{}^{(3)}R = - \frac{6K}{a^2} + \frac{12K\EuScript{R}}{a^2} + 4\nabla ^2 \EuScript{R} , 
\end{equation}
where $ \EuScript{R}=\frac{A}{2} $ is the \textit{curvature perturbation} \cite{ref20}. On the other hand, in order to calculate the exterior curvatures of $ t=const $, we must first obtain the \textit{Weingarten map} (\textit{shape operator}) \cite{ref18} associated with the hypersurface which in coordinate basis may be written as  $ {K^i}_j=\nabla_j n^i $. One can show that
\begin{eqnarray}
{K^i}_j &=& \left(H - \frac{1}{2}HE + \frac{1}{2}\dot A \right){\delta ^i}_j - \Bigg[\frac{1}{a}{H_{kj}}\sigma  - \frac{1}{2}\left( \nabla _j \dot{C}_k + \nabla _k \dot{C}_j \right)\nonumber \\
&+& \frac{1}{2a}\left(\nabla _j G_k + \nabla _k G_j \right) - \frac{1}{2}\dot{D}_{kj} \Bigg]\tilde{g}^{ik}.
\end{eqnarray}
Now we can determine the most important exterior curvatures of slice $ t=const $ as follows: 
\begin{description}
\item The \textit{Gauss-Kronecker curvature}  which is defined as
\begin{equation}
\EuScript{K}:=\det\left({K^i}_j  \right)=H^3 + H^2\left(3 \dot{\EuScript{R}} - \frac{3}{2}HE - a\nabla ^2\sigma \right). 
\end{equation}
\item  The \textit{mean curvature} is derived by taking trace from $ {K^i}_j  $
\begin{equation}
\EuScript{M}:=\frac{1}{3}\mbox{tr} \left({K^i}_j  \right)= H + \dot{\EuScript{R}} - \frac{1}{2}HE - \frac{1}{3}a{\nabla ^2}\sigma .   
\end{equation}
\item The \textit{principle} (\textit{normal}) \textit{curvatures} $ \lambda _i \left( i=1,2,3\right)  $ are defined as the eigenvalues of the Weingarten map. By a lengthy calculation one can show
\begin{eqnarray}
&&\lambda _1 = H, \hfill \\
&&\lambda _2 = \lambda _3 = H + \frac{3}{2}\dot{\EuScript{R}} - \frac{3}{4}HE - \frac{1}{2}a\nabla ^2\sigma . 
\end{eqnarray}
\end{description}
It can be seen that although $ \EuScript{K} $ and $ \EuScript{M} $ are exterior curvatures of $ t=const $, in contrary to Ricci curvature, none of them depend on curvature index $ K $. Furthermore,
\begin{eqnarray}
&&\EuScript{M}\left( t, \mathbf{x}\right)=\left[ \EuScript{K}\left( t, \mathbf{x}\right)\right] ^{\frac{1}{3}}=H\left( t, \mathbf{x}\right),\\
&&\lambda _2 \left( t, \mathbf{x}\right)= \lambda _3 \left( t, \mathbf{x}\right)=\frac{1}{2}\left[3H\left( t, \mathbf{x}\right) -H\left( t\right) \right].
\end{eqnarray}
Therefore, by perturbing the FLRW universe, two principle curvatures of the absolute space are changed from $ H\left( t\right) $ into $ \frac{1}{2}\left[ 3H\left( t, \mathbf{x}\right) -H\left( t\right)  \right] $. \\
Now let's turn to the \textit{Gauss theorema egregium} \cite{ref19} for a spacetime which may be written as 
\begin{equation}\label{G}
- \frac{1}{2}{}^{(3)}R + \lambda _1\lambda _2 + \lambda _2 \lambda _3 + \lambda _1 \lambda _3 = 8\pi G\rho .
\end{equation}
Equation (\ref{G}) is the geometrical counterpart of the Einstein's equation involving $ T_{00} $ ($ T $ is the energy-momentum tensor). Indeed, this equation as J. A. Wheeler said, states that the sum of intrinsic and extrinsic curvatures of a spatial slice is a measure of the nongravitational energy density of spacetime\cite{ref19}. Furthermore, the Gauss theorema egregium involves $ - \frac{1}{2}{}^{(3)}R + \lambda _1\lambda _2 + \lambda _2 \lambda _3 + \lambda _1 \lambda _3 $  doesn't depend on the spatial slice choice. By substituting $ {}^{(3)}R $ and $ \lambda_i $s values in equation (\ref{G}) one can find 
\begin{equation}\label{F}
\left\{
\begin{aligned}
&\frac{3K}{a^2} + 3H^2 = 8\pi G\bar{\rho} , \hfill \\
&- 2\nabla ^2\EuScript{R} - \frac{6K}{a^2} \EuScript{R} + 6H\dot{\EuScript{R}} - 3H^2E - 2aH\nabla ^2\sigma  = 8\pi G\delta \rho .\nonumber  
\end{aligned}
\right.
\end{equation}
The first equation is the well-known Friedmann equation, so the second equation may be called \textit{perturbative Friedmann equation} too. \\
Now by combining of equations (\ref{H}) and (\ref{F}) one can find
\begin{equation}\label{LF1}
H^2\left(t,\mathbf{x} \right)=\frac{8\pi G}{3}\rho\left(t,\mathbf{x} \right)-\frac{K}{a^2}+\frac{2}{3}\nabla ^2\EuScript{R}\left(t,\mathbf{x} \right)+\frac{2K}{a^2}\EuScript{R}\left(t,\mathbf{x} \right),
\end{equation}
where $ \rho\left(t,\mathbf{x} \right)=\rho\left(t\right)+\delta\rho\left(t,\mathbf{x} \right) $. This equation may be called the \textit{local Friedmann equation}. On the other hand, by using the perturbative Einstein field equations \cite{ref15}, it can be derived
\begin{eqnarray}
H^2\left(t,\mathbf{x} \right)+\dot{H}\left(t,\mathbf{x} \right)&=&-\frac{4\pi G}{3}\left[\rho \left(t,\mathbf{x} \right) +3p\left(t,\mathbf{x} \right)+a^2\nabla ^2\Pi ^S \left(t,\mathbf{x} \right)\right]\nonumber \\
&+&\frac{1}{6}\nabla ^2 E\left(t,\mathbf{x} \right)+\frac{1}{2}\dot{H}E\left(t,\mathbf{x} \right),\label{R}
\end{eqnarray}
where $ p\left(t,\mathbf{x} \right)=p\left(t\right)+\delta p\left(t,\mathbf{x} \right)  $ and $ \Pi ^S \left(t,\mathbf{x} \right) $ is the scalar anisotropic inertia indicating the imperfectness of the cosmic fluid \cite{ref3,ref15}. Equation (\ref{R}) is very similar to the Raychaudhuri equation in the FLRW universe, i.e.
\begin{displaymath}
H^2+\dot{H}=-\frac{4\pi G}{3}\left(\rho +3p \right). 
\end{displaymath}
 So we may call this equation the \textit{local Raychaudhuri equation}.\\
Now let's suppose 
\begin{equation}\label{LF2}
H^2\left(t,\mathbf{x} \right)=\frac{8\pi G}{3}\rho\left(t,\mathbf{x} \right)-\frac{K+\delta K\left(t,\mathbf{x} \right)}{\left[ a\left(t \right)+\delta a\left(t,\mathbf{x} \right)\right] ^2}.
\end{equation}
Then by comparing equations (\ref{LF1}) and (\ref{LF2}) one can show that
\begin{eqnarray}
&&\delta a\left(t,\mathbf{x}\right)= a\left(t \right) \EuScript{R}\left(t,\mathbf{x} \right),\label{SF1}\\
&&\delta K\left(t,\mathbf{x} \right)=-\frac{2}{3} a^2\left(t\right) \nabla ^2\EuScript{R}\left(t,\mathbf{x} \right).\label{KI1}
\end{eqnarray}
Equations (\ref{SF1}) and (\ref{KI1}) may be written as
\begin{eqnarray}
&a\left(t,\mathbf{x}\right)= a\left(t \right) \left(1+\EuScript{R}\right),\label{SF2}\\
&\mathcal{K}_s\left(t,\mathbf{x} \right)=\mathcal{K}_s\left(t\right)-\frac{2}{3}\nabla^2 \EuScript{R}-\frac{2K}{a^2}\EuScript{R} ,\label{KI2}
\end{eqnarray}
where $ a\left(t,\mathbf{x}\right) $ and $ \mathcal{K}_s\left(t,\mathbf{x} \right) $ are respectively local scalar factor and local sectional curvature of the absolute space. Equation (\ref{KI2}) shows that slice $ t=const $ no longer is space form. Meanwhile, the sectional curvature of $ t=const $ is a scalar random field, so the universe may be thought as a \textit{random manifold}.\\
Clearly, there is a close relation between $ a\left(t,\mathbf{x}\right) $ and $ K\left(t,\mathbf{x} \right) $
\begin{equation}
K\left(t,\mathbf{x} \right)=K-\frac{2}{3}a\nabla^2 a\left(t,\mathbf{x} \right).
\end{equation}
Moreover, equation (\ref{KI1}) reveals more why $ \EuScript{R} $ is called the curvature perturbation. In the shear free (Newtonian) gauge for which $ \sigma\equiv 0 $, One can show
\begin{equation}\label{HA}
H\left(t,\mathbf{x}\right)=\frac{a^\ast \left(t,\mathbf{x}\right)}{a\left(t,\mathbf{x}\right)} ,
\end{equation}
where $ \ast $ here stands for the absolute derivative along the congruence associated with $ n^\mu $ i.e.
\begin{displaymath}
a^\ast\left(t,\mathbf{x}\right)=n^\mu \partial _\mu a\left(t,\mathbf{x}\right).
\end{displaymath}
However, equation (\ref{HA}) doesn't hold generally. The comprehensive definition of the (local) Hubble parameter may be stated in terms of the mean curvature of the spatial slice of the universe. Notice that $ H\left(t,\mathbf{x}\right)\propto\frac{a^\ast \left(t,\mathbf{x}\right)}{a\left(t,\mathbf{x}\right)} $, so the local Hubble parameter represents the rate of expansion of the universe too.\\
Equation (\ref{SF2})  may be generalized as
\begin{equation}
a\left(t,\mathbf{x}\right)= a\left(t \right)\exp\EuScript{R},\label{SF3}
\end{equation}
which is the nonlinear generalization of the curvature perturbation\cite{ref21}. Indeed, if we condone the cosmological principle because of inhomogeneities and anisotropies in the universe and instead suppose the universe is a \textit{globally hyperbolic} manifold, the metric of the universe may be written in the ADM formalism\cite{ref22}
\begin{equation}
ds^2=-\left(N^2-N_iN^i \right) dt^2+2N_idtdx^i+g_{ij}dx^idx^j.
\end{equation}
In general, $ \left(t,\mathbf{x} \right)  $ is not necessarily the comoving coordinates system, however we suppose that $ t $ accords with the cosmic time. Notice that $ N=N \left(t,\mathbf{x} \right) $ , $ N^i=N^i \left(t,\mathbf{x} \right) $ and $ g_{ij}=g_{ij}\left(t,\mathbf{x} \right)  $ where $ N^i $ and $ g_{ij} $ are respectively a spatial vector and a spatial tensor on the leaves of the spacetime. Now we suppose
\begin{equation}
g_{ij} \left(t,\mathbf{x} \right):=\breve{a} \left(t,\mathbf{x} \right)\breve{g}_{ij} \left(t,\mathbf{x} \right),
\end{equation}
where
\begin{equation}
\breve{g}_{ij} \left(t,\mathbf{x} \right):=\left\lbrace \tilde{g}\exp{\left[ \tilde{g}^{-1}\left(\nabla C+\left( \nabla C\right)^T +D  \right)\right]}\right\rbrace  _{ij}.
\end{equation}
Here $ \tilde{g} $, $ \nabla C $ and $ D $ are matrices respectively corresponding to $ \tilde{g}_{ij}=\delta _{ij} + \frac{Kx^i x^j}{1 - K{\bf{x}}^2 } $,  $ \nabla_i C_j $ and $ D_{ij} $. Moreover, superscript $ T $ stands for the matrix transpose. We suppose $ \mbox{tr}\left(\tilde{g}^{-1}D \right) =\mbox{tr}\left(\tilde{g}^{-1}\nabla C \right) \equiv 0 $ or $ \tilde{g}^{ij}D_{ij}=\tilde{g}^{ij}\nabla_iC_j\equiv 0 $ too, so $ \det \breve{g}=\det \tilde{g} $. $ C_i=C_i\left( t, \mathbf{x}\right)  $ and $ D_{ij}=D_{ij}\left( t, \mathbf{x}\right) $ may be thought as the nonlinear generalizations of the vorticial perturbation and cosmological gravitational wave respectively. Furthermore, we assume 
\begin{equation}\label{NR}
\breve{a}\left( t, \mathbf{x}\right):=a\left( t\right)\exp\EuScript{R}\left( t, \mathbf{x}\right).
\end{equation}
One can show that by considering $ \EuScript{R} $, $ C $ and $ D $ as perturbative terms,
\begin{equation}
g_{ij}=a^2\left(t \right)\left[\left( 1+2\EuScript{R}\right) \tilde{g}_{ij}+\nabla _i C_j+\nabla _j C_i+D_{ij} \right],  
\end{equation}
which coincides with equation (\ref{SP}) in some gauges with $ B\equiv 0 $ e.g. Newtonian gauge or comoving gauge\cite{ref14}. \\   
From equation (\ref{SF2}) we can deduce
\begin{equation}
\langle a\left( t, \mathbf{x}\right)\rangle =a\left( t\right).  
\end{equation}
By contrast, for $ \breve{a}\left( t, \mathbf{x}\right) $ we have
\begin{equation}
\ll \breve{a}\left( t, \mathbf{x}\right)\gg =\exp \langle \ln \breve{a}\left( t, \mathbf{x}\right)\rangle =a\left( t\right),
\end{equation}
where $ \ll \gg $ stands for the \textit{geometric mean}. thus  
\begin{equation}
\langle \breve{a}\left( t, \mathbf{x}\right)\rangle\geq \langle a\left( t, \mathbf{x}\right)\rangle .
\end{equation}
Notice that geometric mean is never greater than the arithmetic mean. 
\section{Conclusion}
In this article we answered how important is the curvature index in the evolution of the universe and found that in inflation, curvature -if it is nonzero- plays a crucial role which affects on the dynamics and causal structure of the inflationary epoch. We also investigated non-intrinsic properties of the spatial slice of linearly perturbed FLRW universe and showed that the Gauss theorema egregium accords with the Friedmann equation. Finally, we discuss about local curvature index and local scale factor of the universe.



\end{document}